\documentclass[12pt]{article}

\usepackage{amssymb}
\usepackage{amsmath}
\usepackage{amscd}
\usepackage{latexsym}
\usepackage{graphicx}

\usepackage{cite}

\newcommand{\be}{\begin{equation}}
\newcommand{\ee}{\end{equation}}

\newcommand{\dlt}{\delta}
\newcommand{\prt}{\partial}
\newcommand{\br}{{\bf r}}
\newcommand{\bk}{{\bf k}}

\newcommand{\ba}{{\bf a}}
\newcommand{\bp}{{\bf p}}
\newcommand{\bu}{{\bf u}}

\newcommand{\vp}{\varphi}
\newcommand{\ep}{\varepsilon}
\newcommand{\al}{\alpha}
\newcommand{\ra}{\rightarrow}

\newcommand{\om}{\omega}
\newcommand{\Om}{\Omega}
\newcommand{\Gm}{\Gamma}
\newcommand{\dgr}{\dagger}
\newcommand{\lbd}{\lambda}

\newcommand{\rgl}{\rangle}
\newcommand{\lgl}{\langle}

\begin{document}

\begin{center}

{\Large{\bf From optical lattices to quantum crystals} \\ [5mm]

V.I. Yukalov$^{1,2}$ }  \\ [3mm]

{\it
$^1$Bogolubov Laboratory of Theoretical Physics, \\
Joint Institute for Nuclear Research, Dubna 141980, Russia \\ [2mm]

$^2$Instituto de Fisica de S\~ao Carlos, Universidade de S\~ao Paulo, \\
CP 369,  S\~ao Carlos 13560-970, S\~ao Paulo, Brazil} \\ [2mm]

{\bf E-mail}: {\it yukalov@theor.jinr.ru} 

\end{center}

\vskip 1cm

\begin{abstract}
Optical lattices can be loaded with atoms which can have strong interactions, 
such that the interaction of atoms at different lattice sites cannot be neglected. 
Moreover, the intersite interactions can be so strong that it can force the atoms 
to form a self-organized lattice, such that exists in crystals. With increasing 
intersite interactions, there can appear several lattice states, including conducting 
optical lattices, insulating optical lattices, delocalized quantum crystals, and 
localized quantum crystals. 
\end{abstract}

\vskip 2mm
{\parindent = 0pt
{\it Keywords}: optical lattices, intersite interactions, collective phonon 
excitations, stability of localized states, quantum crystals }

\newpage

\section{Introduction}

Optical lattices can be created by laser beams and loaded with different atoms 
and molecules \cite{Letokhov_1}. The standard situation is when atoms, or molecules, 
interact with each other through local forces modelled by the delta-function potential 
\cite{Morsch_2,Moseley_3,Yukalov_4,Pethick_5}. With such a short-range potential, 
the interactions of atoms at different lattice cites are negligible, and only onsite 
interactions are taken into account. There exist also long-range potentials, such as 
dipolar, for which it is necessary to take account of intersite interactions 
\cite{Ueda_6,Griesmaier_7,Baranov_8,Baranov_9,Gadway_10,Kurn_11,Yukalov_12,Yukalov_13}.
Usually dipolar potentials are treated as weak, in comparison to local potentials. 
However, if they are sufficiently strong, they can lead to roton instability inducing 
self-organized droplet crystals, which also appear for atoms with some soft-core 
repulsive interactions \cite{Boninsegni_14}.

Actually, it is well known that atoms with sufficiently strong interactions, not 
necessarily of long-range type, can form self-organized periodic structures, that 
is crystals. When quantum properties are important, one has quantum crystals 
\cite{Nosanow_15,Guyer_16,Yukalov_17}. Thus it looks to be clear that under weak 
atomic (molecular) interactions, but in the presence of external periodic fields 
forming a lattice, there is the usual optical lattice, with the lattice parameters 
prescribed by external laser fields. While when intersite interactions are sufficiently 
strong, a self-organized periodic structure can develop. Our aim is to understand what 
happens between these two limiting states, an optical lattice and a quantum crystal.    

It is also well known that optical lattices can house either nonlocalized or localized 
atoms. If the atoms in the lattice are not localized, they can be in a normal state or 
in a superfluid state. Nonlocalized states are often termed conducting. But our aim is 
not to study phase transitions between normal and superfluid nonlocalized states, but 
to distinguish between localized and nonlocalized (conducting) states.

Keeping in mind to find a stability boundary between two phases, without investigating 
the details of the related phase transition itself, it is not compulsory to study both 
the neighbouring phases at both sides of this boundary. But it is sufficient to study 
the system stability from one side of the boundary. This is what is done in the present 
paper by determining the stability boundary from the side of a localized state. The 
main aim is to find out what states, from the point of view of particle localization, 
can arise under the variation of atomic interactions from weak, when the system 
presents a simple optical lattice, to strong, when the lattice structure is caused 
by atomic interactions. That is, the aim is to classify possible states arising, under 
the strengthening of atomic interactions, between the standard optical lattices and 
quantum crystals.          

The system of units is employed, where the Planck and Boltzmann constants are set to 
one.

\section{Localized states}

A system of atoms, with mass $m$, interacting by means of a pair potential $\Phi(\br)$,
is also subject to laser beams creating an optical lattice with a periodic potential
\be
\label{1}
U_L \left( \br + \frac{\lbd_\al}{2} \right) = U_L(\br) \;   ,
\ee
with the lattice spacing $\lbd_\al/2$ in the direction labeled by $\al=1,2,\ldots,d$,
where $d$ is the space dimensionality. The typical lattice potential is
\be
\label{2}
 U_L(\br) = \sum_{\al=1}^d U_\al\sin^2 \left( k_0^\al r_\al \right) \;  ,
\ee
where $k_0^\al = 2\pi/\lbd_\al$. 

The system Hamiltonian is
\be
\label{3}
 \hat H = \int \psi^\dgr(\br) \hat H_L(\br) \psi(\br) \; d\br \; + \;
\frac{1}{2} \int \psi^\dgr(\br) \psi^\dgr(\br') 
\Phi(\br-\br') \psi(\br')\psi(\br) \; d\br d\br' \;   ,
\ee
where
\be
\label{4}
 \hat H_L(\br) = -\;\frac{\nabla^2}{2m} + U_L(\br)
\ee
is the lattice Hamiltonian. 

Field operators obey either Bose or Fermi statistics, which for a well localized 
state is of no importance. Keeping in mind a localized state, the field operators 
can be expanded over well localized Wannier functions \cite{Marzari_18}
\be
\label{5}
 \psi^\dgr(\br) = \sum_{j=1}^{N_L} \hat c_j w(\br -\br_j) \; ,
\ee
where the single-band approximation is assumed and $N_L$ is the number of sites. 

Substituting expansion (\ref{5}) into the Hamiltonian gives the form \cite{Yukalov_18}
\be
\label{6}
 \hat H = - \sum_{i\neq j} J_{ij} \hat c_i^\dgr \hat c_j \; + \;
\sum_j \left( \frac{\bp_j^2}{2m} + U_L\right) \hat c^\dgr_j \hat c_j \; + \;
\frac{1}{2} \sum_j U_{jj} \hat c^\dgr_j \hat c^\dgr_j \hat c_j \hat c_j \; + \;
\frac{1}{2} \sum_{i\neq j} U_{ij} \hat c_i^\dgr \hat c^\dgr_j \hat c_j \hat c_i \; ,
\ee
in which the notation is used for the tunneling matrix
\be
\label{7}  
 J_{ij} = - \int w^*(\br-\br_i) \hat H_L(\br) w(\br-\br_j)\; d\br 
\qquad 
( i \neq j) \;  ,
\ee
the matrix element for the momentum squared
\be
\label{8}
\bp_j^2 = \int w^*(\br-\br_j) (- \nabla^2 ) w(\br-\br_j)\; d\br \;  ,
\ee
the lattice Hamiltonian parameter
\be
\label{9}
 U_L = \int |\; w(\br) \; |^2 \; U_L(\br) \; d\br \; ,
\ee
and the interaction matrix
\be
\label{10}
 U_{ij} = \int |\; w(\br-\br_i) \; |^2 \;
\Phi(\br - \br') \; |\; w(\br'-\br_j) \; |^2 \; d\br d\br' \;  .
\ee
  
The density of atoms is
\be
\label{11}
 \rho(\br) = \lgl \; \hat\psi^\dgr(\br) \hat\psi(\br) \; \rgl =
\sum_{ij} \; \lgl \; \hat c^\dgr_i \hat c_j \; \rgl \; 
w^*(\br-\br_i )  w(\br-\br_j) \; .
\ee
We keep in mind a well localized lattice, assuming the use of well localized Wannier
functions \cite{Marzari_18}. Under the assumption of strong localization, the intersite 
tunneling of atoms can be neglected setting
\be
\label{12}
\hat c^\dgr_i \hat c_j = \nu \dlt_{ij} 
\qquad 
\left( \nu \equiv \frac{N}{N_L} \right) \; ,
\ee
with $\nu$ filling factor. Then the density of atoms (\ref{11}) reads
\be
\label{13}
 \rho(\br) = \nu \sum_j  |\; w(\br-\br_j) \; |^2 \; .
\ee

\section{Energy of atoms}

Atomic energy levels can be defined \cite{Bonch_19,Yukalov_20} as the eigenvalues 
of the eigenproblem
\be
\label{14}
\int H(\br,\br',\om) \vp_{nk}(\br') \; d\br' = E_{nk} \vp_{nk}(\br) 
\ee
for the effective Hamiltonian
\be
\label{15}
 H(\br,\br',\om) = -\; \frac{\nabla^2}{2m} \; \dlt(\br-\br') + 
\Sigma(\br,\br',\om) \; ,
\ee
in which $\Sigma$ is self-energy. For well localized particles, the self-energy can 
be taken \cite{Nosanow_15,Guyer_16,Yukalov_17} in the Hartree approximation
\be
\label{16}
 \Sigma(\br,\br',\om) = [\; U_L(\br) + V_H(\br) \; ] \dlt(\br-\br') \; ,
\ee
with the Hartree potential
\be
\label{17}
  V_H(\br) = \int \Phi(\br - \br') \rho(\br') \; d\br ' \;  .
\ee
Then eigenproblem (\ref{14}) reduces to the equation
\be
\label{18}
 \hat H(\br) \vp_{nk}(\br) = E_{nk} \vp_{nk}(\br) \; ,
\ee
with the periodic Hamiltonian
\be
\label{19}
 \hat H(\br)  = \hat H_L(\br)  +  V_H(\br) =  \hat H(\br+ \br_j) \; ,
\ee
whose eigenfunctions are, clearly, Bloch functions. 

It is useful to mention that, generally, Wannier functions are not the eigenfunctions 
of Hamiltonian (\ref{19}). To see this, it is sufficient to expand the Bloch function 
over
Wannier functions,
$$
 \vp_{nk}(\br) = \frac{1}{\sqrt{N_L}} \; 
\sum_j w_n(\br - \br_j) e^{i\bk\cdot\br_j} \; , 
$$
which transforms eigenproblem (\ref{18}) into the equation 
\be
\label{20}
  \hat H(\br) w_n(\br - \br_j) = \sum_i E_{ij}^n \; w_n(\br - \br_i) \; ,
\ee
in which
\be
\label{21}
 E_{ij}^n \equiv \frac{1}{N_L} \sum_k E_{nk} e^{i\bk\cdot\br_{ij} } 
\qquad
(\br_{ij} \equiv \br_i - \br_j ) \;  .
\ee

The eigenfunctions of Hamiltonian (\ref{19}) are not separate Wannier functions but 
the lattice sums
\be
\label{22}
\psi_n(\br) = \sum_j w_n(\br - \br_j )
\ee
that are the eigenfunctions satisfying the eigenproblem
\be
\label{23}
 \hat H(\br)\psi_n(\br) = E_n \psi_n(\br) \;  ,
\ee
with the eigenvalues
\be
\label{24}
  E_n \equiv \sum_j E_{ij}^n = \lim_{k \ra 0} E_{nk} \; .
\ee

For well localized Wannier functions, we have
\be
\label{25}
 \psi_n(\br) \approx w_n(\br - \br_j ) \qquad ( \br \approx \br_j) \; .
\ee
In that case, Wannier functions play the role of approximate eigenfunctions of 
Hamiltonian (\ref{19}). For the lowest band, we get
\be
\label{26}
  \hat H(\br) w(\br) = E w(\br) \qquad (\br \approx 0 ) \; .
\ee

The solution to equation (\ref{26}) can be found by means of optimized perturbation
theory \cite{Yukalov_21}. For the initial approximation, it is natural to take a 
harmonic Hamiltonian
\be
\label{27}
\hat H_0(\br) = -\; \frac{\nabla^2}{2m} + U_0(\br) \; ,   
\ee 
with the potential
\be
\label{28}
U_0(\br) = u_0 + \sum_\al \frac{m}{2} \; \ep_\al^2 r_\al^2 \;  ,
\ee
which yields the Gaussian wave function 
\be
\label{29}
 w(\br) = \prod_\al \left( \frac{1}{\pi l_\al^2} \right)^{1/4} 
\exp \left( -\; \frac{r_\al^2}{2l_\al^2} \right) \;  ,
\ee
where $l_\al \equiv 1/\sqrt{m\ep_\al}$. The frequency $\ep_\al$ is the control 
parameter to be found from optimization conditions \cite{Yukalov_21}. 
 
The energy, corresponding to Hamiltonian (\ref{27}), reads
\be
\label{30}
E_0 = u_0 + \frac{1}{2} \sum_\al \ep_\al \;  .   
\ee
And to first order of perturbation theory, the eigenvalue of equation (\ref{26}) is
\be
\label{31}
 E_1 = \int w(\br) \hat H_1(\br) w(\br) \; d\br \; ,
\ee
where
$$
\hat H_1(\br) = - \; \frac{\nabla^2}{2m} + U_1(\br) \; , \qquad
U_1(\br) = U_L(\br) + V_1(\br) \; ,
$$
$$
 V_1(\br) = \int \Phi(\br-\br') \rho_0(\br') \; d\br' \; , \qquad
\rho_0(\br) = \nu \sum_j w^2(\br - \br_j) \; .
$$
As an optimization condition, it is convenient to take the equality
\be
\label{32}
 E_1 - E_0 = 0 \; .
\ee

For small $r$, the lattice potential (\ref{2}) can be written as 
\be
\label{33}
U_L(\br) \simeq \sum_\al \frac{m}{2} \; \om_\al^2 r_\al^2 
\qquad 
(\br \approx 0 ) \; ,
\ee
with the frequency
\be
\label{34}
 \om_\al = 2 \sqrt{ E_R^\al U_\al }  \qquad 
\left( E_R^\al \equiv \frac{(k_0^\al)^2}{2m} \right) \;  .
\ee
Then for energy (\ref{31}) it follows
\be
\label{35}
 E_1 = u_0 + \frac{1}{4} \sum_\al \left( \ep_\al + 
\frac{\om_\al^2}{\ep_\al} + \frac{\Om_\al^2}{\ep_\al} \right) \; ,
\ee
where
\be
\label{36}
 \Om_\al \equiv \left[ \; \frac{2\nu}{m} 
\sum_j \frac{\prt^2\Phi(\br_j)}{\prt r_j^\al \prt r_j^\al} \; \right]^{1/2} \; .
\ee
Setting 
\be
\label{37}
 u_0 = \nu \sum_j \Phi(\br_j) \; ,
\ee
we find the effective frequency of atomic oscillations
\be
\label{38}  
 \ep_\al = \sqrt{\om_\al^2+\Om_\al^2} \; .
\ee

As is seen, the effective frequency of atomic oscillations is formed both, by the 
lattice potential and by direct atomic interactions. When the atomic interaction 
potential grows from zero to a strong $\Phi({\bf r})$, the effective frequency 
increases from $\om_\al$, created by the optical lattice, to $\Om_\al$, due to 
atomic interactions. Let us mention that in the case of well localized atoms, the 
intersite tunneling is suppressed, which is evident from the expression for the 
tunneling parameter
$$
 J_{ij} =\sum_\al \left\{ \frac{\ep_\al}{8} 
\left[ \left( \frac{r_{ij}^\al}{l_\al} \right)^2 - 2 \right] - U_\al \right\}
\; \exp\left\{ - \; \frac{1}{4} 
\sum_\al \left(\frac{r_{ij}^\al}{l_\al} \right)^2 \right\} \;  .
$$

\section{Phonon excitations}

The lowest atomic band corresponds to the ground-state energy. Above this energy 
there exist collective atomic excitations represented by phonon degrees of freedom. 
Phonon variables can be introduced as it has been done, e.g., in Refs. 
\cite{Yukalov_22,Yukalov_23,Yukalov_24}. The vector ${\bf r}$ is treated as an 
operator written in the form
\be
\label{39}
  \br_j = \ba_j + \bu_j \; ,
\ee
in which ${\bf a}_j$ is an equilibrium position defining an effective lattice site, 
while ${\bf u}_j$ is a deviation from the site. Thus the averages
\be
\label{40}
\ba_j = \lgl \; \br_j \; \rgl \; , \qquad \lgl \; \bu_j \; \rgl = 0 \; ,
\ee 
are assumed to be valid by definition. 

Note that the actual positions of the effective lattice sites ${\bf a}_j$ are given
by the minimization of a thermodynamic potential. This implies that the parameters 
of the effective lattice are defined by the joint action of the imposed optical 
lattice and by atomic interactions.  

An important quantity is the mean-square atomic deviation 
\be
\label{41}
 r_0^2 \equiv \sum_{\al=1}^d \; \lgl \; u_j^\al u_j^\al \; \rgl \; .
\ee
In the Debye approximation, we find \cite{Yukalov_23,Yukalov_24} 
\be
\label{42}
 r_0^2 = \frac{\nu d}{2m\rho} \int_\mathbb{B} \; \frac{1}{\om_k}\;
\coth\left( \frac{\om_k}{2T} \right) \frac{d\bk}{(2\pi)^d} \;  ,
\ee
where the integration is over the Brillouin zone and the phonon frequency is defined 
by the expression
\be
\label{43}
 \om_k^2 = \frac{4\nu}{m} \; D \;
\sum_{\al=1}^d \sin^2\left( \frac{k_\al a}{2}\right) \;  ,
\ee
with the notation for the dynamic parameter
\be
\label{44}
 D = \frac{\nu^2}{d} 
\sum_\al \; \frac{\prt^2 U(\ba)}{\prt a^\al \prt a^\al} \; ,
\ee
in which
$$
 U(\ba_{ij}) = \int |\; w(\br-\ba_i) \; |^2 \; \Phi(\br-\br') \;
|\; w(\br'-\ba_j) \; |^2  \; d\br d\br'
$$
and where the nearest-neighbour approximation is used, with $a$ being the 
nearest-neighbour distance.

The long-wave limit gives the phonon spectrum
\be
\label{45}
\om_k \simeq c_0 k \qquad 
\left( k^2 \equiv \sum_{\al=1}^d k_\al^2 \right) \; ,
\ee
with the sound velocity
\be
\label{46}
 c_0 = \sqrt{\frac{\nu}{m}\; D a^2 } \;  .
\ee
In the Debye approximation, the phonon spectrum is limited by the Debye wave 
vector
$$
k_D =  \frac{\sqrt{4\pi}}{a} \; \left[ \; \frac{d}{2} \; 
\Gm\left( \frac{d}{2} \right) \; \right]^{1/d} \; .
$$
 
The phonon excitations do not destroy localized states, provided the Lindemann 
criterion of stability holds true \cite{Lindemann_25}. According to this criterion, 
a localized state can be stable, if the mean-square deviation is smaller than half 
of the nearest-neighbour distance,
\be
\label{47}
 \frac{r_0}{a} < \frac{1}{2} \; .
\ee 
Otherwise, the localized state is not stable. In the Debye approximation, the 
stability criterion reads
\be
\label{48}
  12 \; \frac{E_K}{T_D} \int_0^1 x^{d-2} 
\coth\left( \frac{T_D}{2T} \; x \right) \; dx ~ < ~ 1 \; ,
\ee
where the Debye temperature is
\be
\label{49}
 T_D = \sqrt{4\pi \; \frac{\nu D}{m} } \; \left[ \;
\frac{d}{2} \; \Gm\left( \frac{d}{2} \right) \; \right]^{1/d}  
\ee
and $E_K$ is a characteristic kinetic energy
\be
\label{50}
 E_K \equiv \frac{1}{2ma^2} \; .
\ee

At zero temperature, the localized state is stable under the condition
\be
\label{51}
 \frac{T_D}{E_K} ~ > ~ \frac{4d^2}{d-1} \qquad ( T = 0 ) \; .
\ee
This means that a one-dimensional crystal is not stable. The criterion for a 
two-dimensional crystal is
\be
\label{52}
 \frac{T_D}{E_K} ~ > ~ 16 \qquad (d = 2 , ~ T = 0 ) \; .
\ee 
And a three-dimensional crystal at zero temperature can be stable, if
\be
\label{53}
 \frac{T_D}{E_K} ~ > ~ 18 \qquad (d = 3 , ~ T = 0 ) \; .
\ee
  
At high temperatures, the criterion becomes
\be
\label{54}
\frac{T_D}{E_K} ~ > ~ \frac{8d^2}{d-2} \qquad ( T \gg T_D ) \; .
\ee
Hence only a three-dimensional crystal can be stable, provided that
\be
\label{55}
\frac{T_D}{E_K} ~ > ~ 72 \qquad (d = 3 , ~ T \gg T_D ) \; .
\ee

For a three-dimensional system, the oscillation frequency (\ref{36}), caused by 
atomic interactions, is close to the Debye temperature (\ref{49}) and the recoil 
energy $E_R$ is close to the characteristic kinetic energy $E_K$. The system state 
depends of the relation between three main parameters, the frequency $\om_\al$, 
defined by the optical lattice, the characteristic kinetic energy $E_K$, and the 
Debye temperature $T_D$. Summarizing the available information and assuming that 
the frequencies $\om_\al$ are close to each other, we find the following 
classification of possible states.  

If the imposed optical lattice is shallow, such that $\om_\al\ll E_K$, then, 
depending on the relation between the other parameters, there can exist the following 
states. If $0\leq T_D\ll \om_\al$, the system forms a conducting optical lattice. 
When $\om_\al\ll T_D\ll E_K$, a delocalized quantum crystal can arise. For 
$\omega_\alpha \ll E_K \ll T_D$, a localized quantum crystal is stable.   

When the optical lattice is sufficiently deep, so that $\om_\al\gg E_K$, then the 
following states are possible. Here we keep in mind the general situation with an 
arbitrary filling factor and temperature. The case of an integer filling factor and 
zero temperature is special, requiring to take account of the onsite interactions  
\cite{Morsch_2,Moseley_3,Yukalov_4,Pethick_5}. If $0<T_D\ll E_K$, we have a conducting 
optical lattice. When $E_K \ll T_D \ll \omega_\alpha$, the optical lattice becomes 
insulating. And for $E_K \ll \omega_\alpha \ll T_D$, a localized quantum crystal 
is formed.   

Concluding, the role of intersite interactions is investigated for a system of 
atoms in an optical lattice. In the absence of the intersite interactions, all 
properties of the system are governed by the optical lattice. But in the presence 
of the intersite interactions, collective phonon excitations can arise, and the 
system state can be strongly changed. This state is mainly regulated by the relations 
between three parameters, the frequency $\om_\al$, caused by the optical lattice, the 
characteristic kinetic energy $E_K$, that is close to the recoil energy, and the Debye 
temperature $T_D$, whose existence and strength are due to the intersite interactions. 
A three-dimensional system, with an arbitrary filling factor and temperature, can be 
in four different states forming either a conducting optical lattice, or an insulating 
optical lattice, or a delocalized quantum crystal, or a localized quantum crystal. In 
a quantum crystal there also exists a lattice. But the difference between an optical 
lattice and a crystalline lattice is in their origins. An optical lattice is created 
by external laser beams prescribing the lattice period and depth. In contrast, a 
crystalline lattice is formed self-consistently, due to atomic interactions, providing 
a state realizing a minimum of thermodynamic potential. 

\section*{Acknowledgment}

The authors is grateful to E.P. Yukalova for discussions and advice.


\vskip 2cm


\begin{thebibliography}{99}
\bibitem{Letokhov_1}
Letokhov V  2007
{\it Laser Control of Atoms and Molecules} (New York: Oxford University)

\bibitem{Morsch_2}
Morsch O and Oberthaler M  2006
{\it Rev. Mod. Phys.} {\bf 78} 179 

\bibitem{Moseley_3}
Moseley C, Fialko O and Ziegler K  2008
{\it Ann. Phys. (Berlin)} {\bf 17} 561 

\bibitem{Yukalov_4}
Yukalov V I  2009  
{\it Laser Phys.} {\bf 19} 1 

\bibitem{Pethick_5}
Pethick C J and Smith H  2008  
{\it Bose-Einstein Condensation in Dilute Gases} 
(Cambridge: Cambridge University) 

\bibitem{Ueda_6}
Ueda M  2010 
{\it Fundamentals and New Frontiers of Bose-Einstein Condensation} 
(Singapore: World Scientific)

\bibitem{Griesmaier_7}
Griesmaier A  2007 
{\it J. Phys. B} {\bf 40} R91 

\bibitem{Baranov_8}
Baranov M A  2008 
{\it Phys. Rep.} {\bf 464} 71 

\bibitem{Baranov_9}
Baranov M A, Dalmonte M, Pupillo G, and Zoller P  2012 
{\it Chem. Rev.} {\bf 112}, 5012 

\bibitem{Gadway_10}
Gadway B and Yan B  2016 
{\it J. Phys. B} {\bf 49} 152002 

\bibitem{Kurn_11}
Stamper-Kurn D M and Ueda M  2013 
{\it Rev. Mod. Phys.} {\bf 85} 1191 

\bibitem{Yukalov_12}
Yukalov V I and Yukalova E P  2016 
{\it Laser Phys.} {\bf 26} 045501  

\bibitem{Yukalov_13}
Yukalov V I 2018 
{\it Laser Phys.} {\bf 28} 053001 

\bibitem{Boninsegni_14}
Boninsegni M  2012
{\it J. Low Temp. Phys.} {\bf 168} 137

\bibitem{Nosanow_15}
Nosanow L H  1966
{\it Phys. Rev.} {\bf 146} 120

\bibitem{Guyer_16}
Guyer R  1969
{\it Solid State Phys.} {\bf 23} 413 

\bibitem{Yukalov_17}
Yukalov V I and Zubov V I  1983
{\it Fortschr. Phys.} {\bf 31} 627  

\bibitem{Marzari_18}
Marzari N, Mostofi A A, Yates J R, Souza I and Vanderbilt D  2012
{\it Rev. Mod. Phys.} {\bf 84} 1419  

\bibitem{Yukalov_18}
Yukalov V I  2020
{\it Laser Phys.} {\bf 30} 015501

\bibitem{Bonch_19}
Bonch-Bruevich V L and Tyablikov S V  1962
{\it The Green Function Method in Statistical Mechanics}  
(Amsterdam: North-Holland)

\bibitem{Yukalov_20}
Yukalov V I  1998
{\it Statistical Green's Functions} (Kingston: Queen's University)

\bibitem{Yukalov_21}
Yukalov V I  2019
{\it Phys. Part. Nucl.} {\bf 50} 141

\bibitem{Yukalov_22}
Yukalov V I  2010
{\it Symmetry} {\bf 2} 40

\bibitem{Yukalov_23}
Yukalov V I and Ziegler K  2015
{\it Phys. Rev. A} {\bf 91} 023628 

\bibitem{Yukalov_24}
Yukalov V I and Ziegler K  2016
{\it J. Phys. Conf. Ser.} {\bf 691} 012014 

\bibitem{Lindemann_25}
Lindemann F A  1910
{\it Z. Phys.} {\bf 11} 609 
\end{thebibliography}
\end{document}